# Transcripts per million ratio: applying distribution-aware normalisation over the popular TPM method

Hilbert Lam Yuen In*, Robbe Pincket
* corresponding author

**ABSTRACT**

Current popular methods in literature of RNA sequencing normalisation do not account for gene length when compared across samples, whilst adjusting for count biases in the data. This creates a gap in the normalisation as bigger genes in RNA sequencing accumulate more reads due to shotgun sequencing methods. As a result, the proportions of these reads inter-sample are not properly accounted for in current normalisation methods. Alternatively, methods which account for gene length do not account for the pan-sample biases in the data by accounting for a central read average. Thus, in order to fill in the gap in the literature, we propose a novel method of Transcripts Per Million Ratio and its relatives in RNA-sequencing differential expression normalisation that can be used in different conditions, which takes into account the gene length as well as relative expression in normalisation.

**INTRODUCTION**

RNA-sequencing is a widely deployed method for studies to classify differentially expressed (DE) genes (1). Due to the fundamental nature of sequencing, high-throughput data measuring up to terabytes per run can be generated (2), in which deciphering through what is actual signal or not can prove challenging. This calls for proper normalisation of this data to account for factors (such as experimental variability, GC content, sequencing depth, and gene length) that prevent direct comparison of raw data (1) and could have led to some shifts not being detected in previous research owing to lack of proper normalisation techniques (1). By using proper normalisation, it can be better ensured that the biological signal of expression of genes under different conditions can be properly examined, and to reduce type I and type II errors that can significantly affect downstream analysis (1).

Currently available, widely used normalisation methods such as DESeq2 (3) and EdgeR (Trimmed Mean of Means, TMM) (4) do not first account for gene length before the main step of normalisation. This is a critical step in normalisation as it is necessary to correctly rank gene expression levels within a sample as longer genes naturally "accumulate" more reads when counted (15), which can affect gene proportions and thus, normalisations in subsequent normalisations due to the fundamental mathematical methods employed in DESeq2 and TMM. . Although it is in theory possible to use TPM post-DESeq2/TMM normalisation on the "pseudo-counts", this is hardly used in practice, and gene length is only taken into account after the highly crucial DESeq/TMM normalisation steps. Therefore, there is a need to include normalisation methods that take into account fundamental assumptions of DE sequencing but also account for the gene length first.

Furthermore, TPM has been shown to be one of the highest performers (5) compared to other normalisation methods, and does not introduce "new or unwanted structure" to the data (5), and performs the second best reduction in residual variability just before log2 normalisation (5).

There has been much debate (1, 6, 7) over the normalisation methods to be employed for this purpose. Transcripts Per Million (TPM), as first defined by (8), has been commonly cited as a normalisation method which does not perform batch-adjustment or "batch-effect" (5). TPM normalises all the reads within a run so that the sum of all the reads would be exactly 1,000,000. In perspective, TPM puts every read count as a proportion of all the other reads which are mapped to in the run.

However, transcriptome-profiling strictly using TPM has been misused unintentionally by researchers to investigate differentially expressed genes, as stated by Zhao *et al.*, 2020. This is because TPM is easily influenced by sample preparation protocols and should not be used to compare between different tissue parts or even between different parts of the same cell (i.e. cell mitochondria/mtRNA vs nuclear RNA). Zhao *et al.*, 2020 also finds that the strandedness of RNA sequencing samples can greatly potentially bias TPM results if misused for DE analysis. This is mostly because TPM does not normalise for batches of samples or conditions of samples.

In spite of this, TPM remains a valuable method in understanding the proportion of genes (5), and it increases the biological variation and reduces residual variability as demonstrated by (5). Fundamentally, TPM serves as an easy way to understand gene expression profiles as a ratio which can be more easily interpreted.

Therefore, we attempt to address the issues which are stated by Zhao *et al.*, 2020 in TPM normalisation in which TPM does not assume that most genes are not DE, DE and non-DE genes behave similarly and that for every upregulated gene, and there is approximately one downregulated gene (14), whilst still creating a robust method that normalises read counts to a ratio similar to that of the original TPM method.

METHODS

We propose three novel methods in normalising RNA Sequencing data. The Transcripts Per Million Ratio (TPMR) method, which uses a geometric mean of gene ratios that cluster around the 50th percentile of a dataset as a basis for normalisation, Transcripts Per Million Median (TPMM), which adjusts for the skewness in individual sample datasets by setting all the medians in different conditions to be identical so as to account for kurtosis, and Transcripts Per Million Ratio 2 (TPMR2), which combines both the TPMR and TPMM methods.

Parameters in the TPMR and TPMR2 methods have an α variable, which accounts for an estimated proportion of genes which do not change in expression across different samples. This follows with a fundamental assumption proposed by Zhao *et al*., 2020 of balanced expression changes.

Moreover, in all these methods, the original TPM method is initially applied, meaning that a relative fold change to the individual experiment is first conducted, and gene length is accounted for in the final normalised per million counts.

**Mathematical definitions used**

Let $K$ be the set of all conditions
Let $J$ be the set of all samples
Let $n$ be the number of genes
Let $k$ be a condition
Let $j$ be a sample
Let $i$ be a gene
Let $x_i$ be the raw read count for gene $i$

$$(J = \bigcup_{k \in K} k) \qquad (1)$$

**Transcripts Per Million (TPM)**

$$a_i = \frac{x_i \cdot 10^3}{10^6} \qquad (2)$$

$$TPM_i = a_i \cdot \frac{1}{\sum_{i=1}^{n} a_i} \tag{3}$$

**Transcripts Per Million Ratio (TPMR)**

First, we define α = 10. α is a number between 0 and 100, it represents the percentage of genes to be delimited as "housekeeping" genes.

$$\beta_{ik} = \underset{j \in k}{\text{mean}}(TPM_{ij}) \tag{4}$$

Here $\beta_{ik}$ represents the average (arithmetic mean) TPM for gene *i* in condition *k*. The assumption is that between technical replicates (i.e. same condition, different samples), the read count follows a normal distribution. Therefore the arithmetic mean is appropriate for this case.

$$\gamma_{\frac{k_2}{k_1}} = \left\{ \frac{\beta_{1k_2}}{\beta_{1k_1}} \cdots \frac{\beta_{nk_2}}{\beta_{nk_1}} \right\} \tag{5}$$

We then divide gene-wise the average TPM for each condition, to form a set γ, as shown above. This set shows how much upregulation/downregulation there is between the same gene across different conditions. The result of this normalisation makes γ a log-normal distributed set.

$$\tau_{\frac{k_2}{k_1}} = \text{geometric mean}\left(\left\{Q_{50-0.5\alpha\%}\left(\gamma_{\frac{k_2}{k_1}}\right) \cdots Q_{50+0.5\alpha\%}\left(\gamma_{\frac{k_2}{k_1}}\right)\right\}\right) \quad (6)$$

$$\tau_{\frac{k_i}{k_i}} = 1 \quad (7)$$

$$\tau_{\frac{k_2}{k_1}} \cdot \tau_{\frac{k_1}{k_2}} = 1 \quad (8)$$

Assuming α = 10, τ would be a set of all geometric mean ratios of the 45th percentile to 55th percentile. τ as a result would be equal 1 if the same condition is divided by itself. Due to the log-normal distribution of reads and the fact that the division of two log-normal distributions gives a new log-normal distribution, the geometric mean in this case will give a better average as compared to an arithmetic mean, as the geometric mean accounts for outliers better in high kurtosis data (as in the case of RNA sequencing data) and thus is a better account for central tendency.

$$\delta_{k_i} = \underset{k_m \in K}{\text{geometric mean}}\left(\tau_{\frac{k_m}{k_1}}\right) \quad (9)$$

Here we get a geometric mean of all the ratios across the board for all conditions (including itself). We repeat the calculation of δ for all combinations of the set. δ$_k$ is the normalisation factor for condition k.

$$TPMR_{ij} = TPM_{ij} \cdot \delta_{K(j)} \quad (10)$$

Finally, we multiply the original TPM with δ$_k$ for condition k to get the TPMR.

The TPMR across samples can then be compared using standard parametric/non-parametric statistical methods, whichever is deemed appropriate, for DE analysis.

**Transcripts per million median (TPMM)**

TPMM normalises the median gene expression of TPM instead of the mean, taking the geometric mean of medians as reference.

$$\theta_j = \text{median}(\{TPM_{1j} \ldots TPM_{nj}\}) \qquad (11)$$

First, we find the median number of reads in the sample. The mean of all TPM reads is always ($10^6$/n), due to the definition of TPM. The usage of the median accounts for the skewness in the distribution which the mean does not account for.

$$\mu = \underset{j \in J}{\text{geometric mean}}(\theta_j) \qquad (12)$$

We then define the geometric mean of all the sample median reads, μ. The rationale for using the geometric mean is that μ is used in the ratio μ/θ, which can not only be seen as a ratio between an average of a mean and a value, but also as an average of ratios.

$$TPMM_{ij} = \frac{\mu}{\theta_j} \cdot TPM_{ij} \qquad (13)$$

Finally, we normalise the TPM by multiplying it, creating a TPMM that is on a similar scale to that of the original TPM, but not *sample-blind* median normalised.

TPM ensures that the *mean* of all genes in each sample is always ($10^6$/n), whereas TPMR, when used cross-sample, ensures that the *median* of all genes in each sample is the same.

**Transcript per million ratio 2 (TPMR2)**

TPMR2 has the same formula as TPMR, but instead of running TPM as a first step, runs TPMM instead, and normalises the TPMM values. TPMR2 is neither *condition-blind* nor *sample-blind*. It normalises for sample-blindness first and ensures that all the medians are closer to each other, before normalising for housekeeping genes. As with TPMR, it requires an α parameter, usually set to α = 10.

$$\beta_{ik} = \operatorname*{mean}_{j \in k}(TPMM_{ij}) \tag{14}$$

$$TPMR2_{ij} = TPMM_{ij} \cdot \delta_{K(j)} \tag{16}$$

**Code**

The aforementioned formulas are coded in Python and are available both as a PIP package and also on GitHub as the package Ribonorma: https://github.com/Chokyotager/Ribonorma

**Reference RNA sequencing experiments**

For reference reads used in the appendix data, we used publicly available data at Recount (9) for analysis at http://bowtie-bio.sourceforge.net/recount/.

Table 1: list of reference reads

| Study | Authors |
|---|---|
| Alternative isoform regulation in human tissue transcriptomes | Wang *et al.*, 2008 |
| Evaluation of statistical methods for normalization and differential expression in mRNA-Seq experiments | Bullard *et al.*, 2010 |

# DISCUSSION

## Application of methods to existing data

The TPMR, TPMM and TPMR2 methods were used on existing *Homo sapiens* data from Wang *et al.*, 2008 and Bullard *et al.*, 2010. This data is shown in the appendix.

In using TPMR normalisation, there is an approximately equal number of upregulated to downregulated genes as shown in Figure 1, which is an issue stated by Zhao *et al.*, 2020.

## Defining housekeeping genes

We define the housekeeping genes in this paper as genes that have a similar expression profile across different samples and/or conditions. The TPMR and TPMR2 normalisation method defines these as the ratioed genes that are near the median between two different conditions.

## Removal of all-zero reads

Removal of all-zero reads across all samples in different conditions is necessary as it can skew the median of the data, as strongly Pareto (12) distributions will shift the median, leading to less ideal housekeeping genes being selected.

## Assumptions used in TPMR, TPMM, and TPMR2

In TPMR, TPMM and TPMR2 normalisations, we assume:

1. Runs under the same condition (i.e. replicates) are biologically identical and any variation follows a normal distribution or Student's t-distribution (TPMR/TPMR2)

2. Genes that are not differentially expressed ("housekeeping genes") are near the median each experimental condition's TPM normalised read counts (TPMR/TPMR2)

3. There is generally the same level of upregulated genes in one condition as there are downregulated genes (TPMR/TPMR2)

4. Housekeeping genes are near the median of each sample's TPM normalised read counts (TPMM/TPMR2)

5. The final scale of the normalisation should be directly comparable to that of the original transcripts per million (TPMR/TPMM/TPMR2)

## Sample, pan-sample and pan-condition blindness

TPM adjusts for individual mapped library sizes in individual samples by normalising the read counts as ratios of one another. TPM adjusts not for the full library size, but only the library which has mapped successfully to the reference - what we deem as effective library size. It is not sample-blind, as it sees the sample as a whole before normalisation.

TPMM adjusts for the skewness of the read distribution across samples of the same condition (i.e. biological replicates of an experiment) by assuming that the median TPM read count is the same throughout the samples. It therefore examines the replicates and normalises them all at once, and is therefore not pan-sample blind.

TPMR adjusts for the ratio of the read count of genes neighbouring the median of ratios in different conditions in equation 3. It is fundamentally similar to median ratio normalisation (MRN), however uses the geometric mean of medially-expressed genes (ratio of which is defined by α). This is as the median may not be fully representative of the distribution, and a geometric mean would represent the skew better of a logarithmic distribution (13), as read counts are logarithmic in nature. Since TPMR does not normalise individually for samples, but instead takes into account all conditions of data which are provided, it is not pan-condition blind.

TPMR2 combines both TPMM and TPMR methods. It first normalises the median expressions between samples, and thus is not pan-sample blind. Subsequently, the values are then normalised through the TPMR method, throughout different conditions presented, thus is also not pan-condition blind.

Since all the metrics which we propose rely on the TPM method, none of them are sample-blind and express the reads as a proportion to that of the effective library size.

**Adjusting least when necessary**

The use of the TPMR ratio adjusts for housekeeping genes in which the percentage of can be defined using the α variable. TPMR2 and TPMM, on the other hand, further normalise for individual sample skewness by ensuring that the median of each sample is the same throughout.

It is recommended to plot a histogram of read count distributions and to check the medians of each sample before employing TPMM and/or TPMR2. If the read medians are generally similar to each other, with no specific extreme skewness in each sample, the authors recommend using the TPMR method for pan-conditions and TPM for single condition pan-samples.

However, if there is strong skewness presented in the data, we would advise the use of TPMR2 for pan-conditions and TPMM for pan-samples instead due to making the medians the same.

TPMM and TPMR2 are generally more "liberal" measures compared to TPMR and tend to show more genes as being DE, and should be used only when the situation is appropriate.

**No one-size-fits-all normalisation**

Although the normalisation methods which are proposed adjust for some of the common issues in comparing TPM alone, we do not suggest it to be a catch-all solution to all RNA-sequencing methods.

Fundamentally, any normalisation method or scheme can be misused. It is of utmost importance that the biological processes of RNA sequencing be fully understood, and proper quality control be performed on all steps of the bioinformatics pipeline before normalisation.

Under very different conditions, mapping methods, QC processes, the signal of the RNA repertoire may significantly change (14). It is therefore prudent to always verify samples and establish robust processing methods, and to use the most suitable normalisation method for respective tasks.

**Pre-normalisation**

We propose the removal of read counts that are 0 to prevent further skewing of the data. As TPMM, TPMR and TPMR2 all require the use of the geometric mean and/or median, the removal of 0 data points which are common in RNA sequencing experiments is fundamental to prevent excessive kurtosis.

**Post-normalisation statistical tests**

We propose the use of non-parametric statistical tests in analysing post-normalised data.

Between two unpaired samples with biological and/or technical replicates, we recommend the use of Welch's t-test or Yuen-Welch's t-test. This is because the variance between two groups may not be the same, as shown in Figure 2 in the appendix.

Between two paired samples with biological and/or technical replicates, we recommend the use of Wilcoxon's signed rank test and the use of the median to compare the fold changes.

Between two unpaired samples in which the data is not approximately normal, we recommend the use of Mann-Whitney's U test.

Lastly, between many different samples, we recommend the use of Kruskal-Wallis H test, followed by Dunn's post-hoc test and two-stage Benjamini, Krieger, & Yekutieli FDR correction to reduce the probability of type I errors (false positives).

Alternatively, Bayesian estimation can be used from the final counts, in which correction will not be necessary. However, this was not performed in our study.

## CONCLUSION

Overall, due to the complicated nature of RNA-sequencing and its exponentially increasing applications, there is, understandably, increasing debate (1, 6, 7) on the most robust measures for RNA-sequencing normalisation and great difficulty in selecting normalisation methods.

We therefore propose the use of TPMR, TPMM and TPMR2 normalisation methods as a benchmark for comparison in RNA-sequencing normalisation, which retains its original similarity and comparability to that of TPM, whilst addressing common issues that present when using TPM to compare across different conditions (14).

## ACKNOWLEDGEMENTS

We thank Dr. Ramasamy Srinivas for reviewing and providing feedback for this work.

## CONFLICTS OF INTEREST

The authors declare no conflict of interest in the production of this work.

# APPENDIX

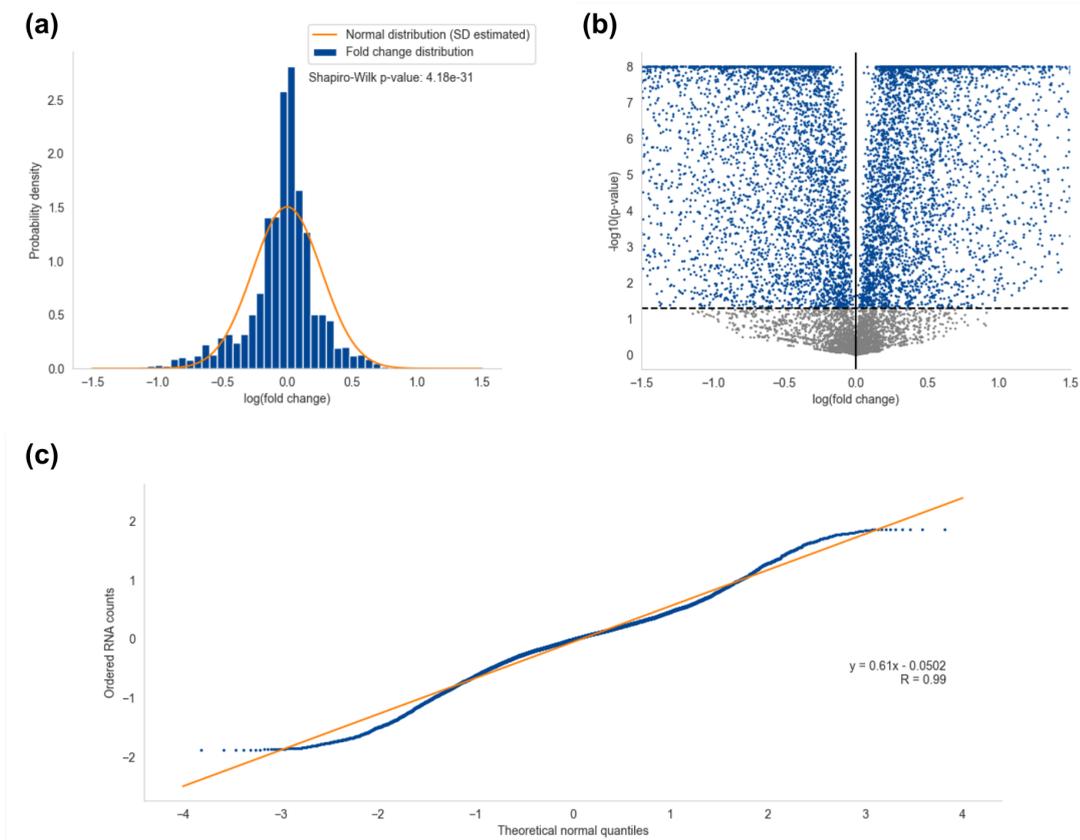

**Figure 1: equal number of upregulated genes as there are downregulated genes when using TPMR normalisation.** RNA sequencing reads from Bullard *et al*., 2010 containing UHR and brain RNA sequencing samples were normalised using TPMR at α = 10, tested using Welch's t-test and plotted in (a) a histogram for all -log(fold change) values between ±1.87 that have a p-value of < 0.05 - followed by a Sharpiro-Wilk's test for normality of the =data - which shows that the data does not follow a strictly normal distribution, requiring more tests, (b) a volcano plot of the -log(p-value) over the log(fold change), with dotted horizontal line indicating a p-value of 0.05, (c) a quantile-quantile plot with a near-perfect Pearson's regression showing all -log(fold change) values between ±1.87.

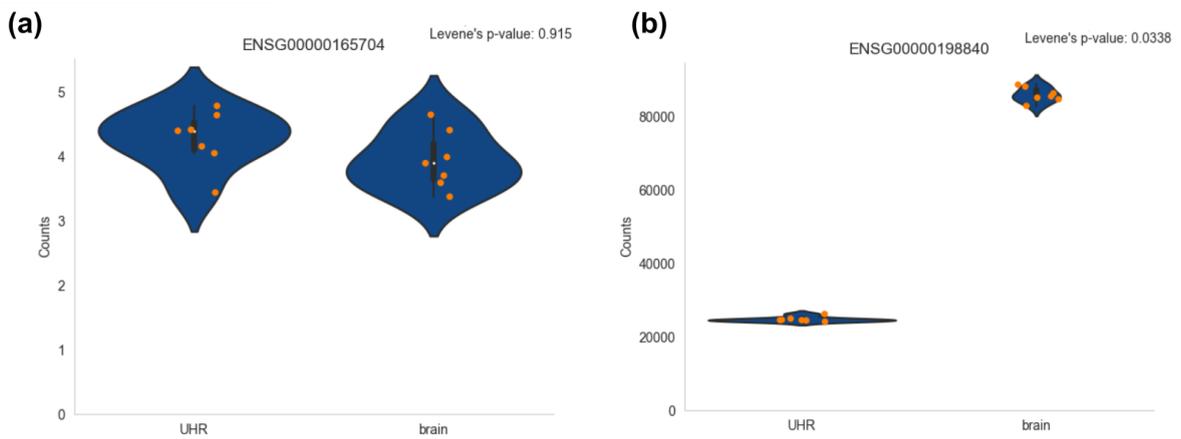

**Figure 2: variances between genes in different phenotypes/conditions can be different in TPMR2 normalisation.** RNA sequencing reads from Bullard *et al.*, 2010 containing UHR and brain RNA sequencing samples were normalised using TPMR2 at α = 10. (a) is the HPRT1 gene, whereas (b) is the MT-ND3 gene. Levene's test of equal variances is used, comparing from the mean. Whereas some genes may have similar variance, other genes have different variances, therefore non-parametric tests are preferred.

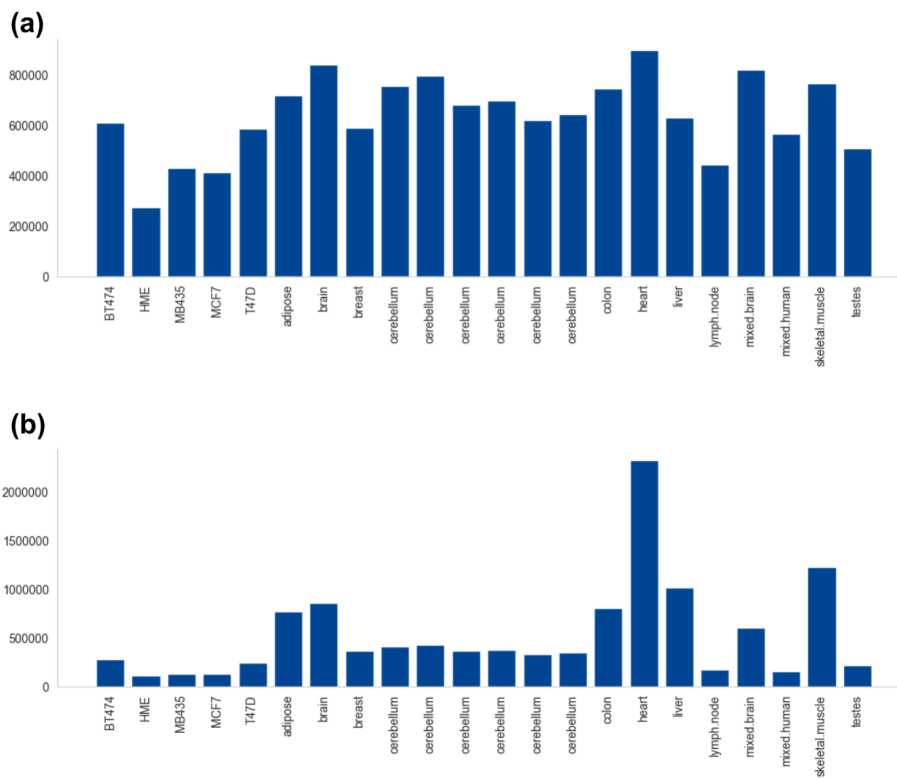

**Figure 3: mtRNA ratios between different tissues are more different in TPMR-normalised counts than in raw TPM.** Counts from Wang *et al.*, 2008 were normalised using TPM in (a) and TPMR at α = 10 in (b) and plotted.